# Closed-form Solutions for an Explicit Modern Ideal Tontine with Bequest Motive

John Dagpunar, Mathematical Sciences, University of Southampton,

University Road, Southampton, SO17 1BJ, United Kingdom

Email: jdagpunar@hotmail.com j.dagpunar@soton.ac.uk

Abstract

*In this paper I extend the work of Bernhardt and Donnelly (2019) dealing with modern explicit tontines, as a way of providing income under a specified bequest motive, from a defined contribution pension pot. A key feature of the present paper is that it relaxes the assumption of fixed proportions invested in tontine and bequest accounts. In making the bequest proportion an additional control function I obtain, hitherto unavailable, closed-form solutions for the fractional consumption rate, wealth, bequest amount, and bequest proportion under a constant relative risk averse utility. I show that the optimal bequest proportion is the product of the optimum fractional consumption rate and an exponentiated bequest parameter. I show that under certain circumstances, such as a very high bequest motive, a life-cycle utility maximisation strategy will necessitate negative mortality credits analogous to a member paying life insurance premiums. Typical scenarios are explored using UK Office of National Statistics life tables.*

## 1. Introduction and summary of results

In a recent paper, Bernhardt and Donnelly (2019) model an explicit modern and ideal tontine with a bequest motive. It is ideal in that it assumes that there are an infinite number of members of the tontine. It is explicit in the sense that a member receives in the interval $(t, t+dt)$ a mortality credit of $\lambda(t)X(t)dt$ where $\lambda(t)$ is an assumed deterministic exponentially increasing mortality rate and $X(t)$ is the size of the pension pot at age $t$. The mortality credit arises through the death of other members of the group in that time interval. When the individual dies his/her pension pot is distributed to other members of the tontine in such a way that the scheme is instantaneously actuarially fair. A distinctive feature of the Bernhardt and Donnelly paper is that it includes a bequest motive. Thus, a proportion $1-\alpha$ of the current pension pot is assigned to a bequest account that does not attract mortality credits, while the remaining proportion $\alpha$ is assigned to the tontine account. On the death of the member, the tontine account is shared out amongst members of the tontine. In contrast, the bequest account passes to the estate of the member. With such a bequest motive, the mortality credit becomes $\alpha\lambda(t)X(t)dt$. The problem is to determine an optimum fractional consumption rate, $c(t)$, expressed as a proportion of the pension pot value (wealth), $X(t)$. At all times the amounts in the tontine and bequest accounts are respectively $\alpha X(t)$ and $(1-\alpha)X(t)$. A proportion $1-w(t)$ of the pension pot is invested in a riskless asset which grows at rate $r$ and the remaining proportion $w(t)$ is invested in a risky asset whose log return in $(t, t+dt)$ is $(\mu - 0.5\sigma^2)dt + \sigma dW(t)$ where $\mu > r$ and $W(t)$ is the standard Wiener process. Transfers between the tontine and bequest accounts must be made continuously to maintain the proportion $\alpha$, but since the same investment strategy is used for tontine and bequest accounts this could be managed by formal declarations as to the current proportion in the bequest account.

The individual chooses $c(t)$ in such a way as to maximize the expected utility over a lifetime. Utility is discounted by a time preference rate $\rho$. Smaller values of $\rho$ value future consumption more highly as might be the practice for persons who anticipate the extra benefit of expenditure in old age on health and social care. Specifically, the utility gained through consumption (withdrawal) in $(t, t+dt)$ is $U[c(t)X(t))]dt$ while the contribution to the individual's utility through death at age $\tau$ say, is $bU[(1-\alpha)X(\tau)]$ where $b \geq 0$ is a parameter that expresses the strength of the bequest motive. The utility function is from the constant relative risk aversion family with a power utility function

$$U(x) = \frac{x^\gamma}{\gamma} \tag{1}$$

where $\gamma \in (-\infty, 1)\backslash 0$ . This has a constant relative risk aversion of $1-\gamma$ . Following the approaches of Yaari (1965), Merton (1969, 1971), Richard (1975) and others, the objective is to find

$$\max_{\{w(s),\alpha(s),c(s):s\geq t\}} \mathbb{E}\left(\int_t^\infty e^{-\int_t^s[\lambda(y)+\rho]dy}\{U[c(s)X(s)] + b\lambda(s)U[(1-\alpha(s))X(s)]\}\,ds\middle|X(t) = x\right) \tag{2}$$

subject to a wealth equation

$$\frac{dX(t)}{X(t)} = \big(r + (\mu - r)w(t) + \alpha(t)\lambda(t) - c(t)\big)dt + \sigma w(t)dW(t). \tag{3}$$

In this paper I extend the work of Bernhardt and Donnelly (2019). In their paper they consider two models; one with a fixed proportion $\alpha$ and a fixed fractional consumption rate, $c$, throughout a member's lifetime; the other with fixed $\alpha$ but variable fractional consumption rate, $c(t)$. The approach in the present paper is to allow both these controls to be optimised dynamically throughout the life-cycle. Optimising the dynamic allocation $\alpha^*(t)$ to the tontine account in this way leads to three key findings. Firstly, the solution is now closed-form, with dynamic proportion invested in the bequest account being $c^*(t)b^{\frac{1}{1-\gamma}}$. Secondly, I am able to give explicit understanding of the role of the bequest parameter, *b*. Specifically, for a life-cycle utility-maximising investor

$$b = \left(\frac{bequest\ amount}{monetary\ consumption\ rate}\right)^{risk\ aversion} \tag{4}$$

An interesting empirical finding is that a member with very low risk aversion will borrow to over-invest in a risky asset. Until old age, such a person will consume (withdraw) little and invest mainly in the tontine account in order to build up longevity credits. It is only in older age that more is switched to the bequest account to reduce the chance on death of a large tontine balance being distributed to tontine members. This surprising feature, evident even for relatively small bequest motives, leads to the possibility of huge bequest amounts in old age, albeit with very small probabilities. The third finding is that for high enough bequest motives and for risk aversions lying in a specified interval, the optimal proportion invested in the tontine account becomes negative while the proportion in the bequest amount exceeds 1. This corresponds to a member effectively making life insurance premium payments.

For convenience, the main results are summarised below. Let

$$\beta = r + \frac{\rho - r}{1-\gamma} - \frac{\gamma}{2}\left(\frac{\mu - r}{\sigma}\right)^2\left(\frac{1}{1-\gamma}\right)^2 \tag{5}$$

and let $s$ denote an age of entry to the tontine. Then when

$$b^{\frac{1}{1-\gamma}}\beta < 1 + \frac{b^{\frac{1}{1-\gamma}}}{\int_s^\infty e^{-\int_s^u[\lambda(y)+\beta]dy}du} \tag{6}$$

the optimal strategy for $t \geq s$ is:

$$c^*(t) = \frac{1}{m(t)} \tag{7}$$

where

$$m(t) = b^{\frac{1}{1-\gamma}} + \left[1 - \beta b^{\frac{1}{1-\gamma}}\right] \int_t^{\infty} e^{-\int_t^u [\lambda(y)+\beta]dy} du \,, \tag{8}$$

$$\alpha^*(t) = 1 - c^*(t) b^{\frac{1}{1-\gamma}} \,, \tag{9}$$

$$w^*(t) = \left(\frac{1}{1-\gamma}\right)\left(\frac{\mu - r}{\sigma^2}\right). \tag{10}$$

The wealth at age $t$, given $X(s) = x_s$ is

$$X(t) = \frac{x_s c^*(s) e^{\left(r-\beta+(\mu-r)w^*-0.5\sigma^2 w^{*2}\right)(t-s)} e^{\sigma w^*[W(t)-W(s)]}}{c^*(t)} . \tag{11}$$

Of note is that $c^*(t)$ , $\alpha^*(t)$ , and $w^*(t)$ are all deterministic, with the latter independent of $t$.

The paper is organised as follows. In section 2, I provide more context and review relevant literature. Section 3 develops the mathematical analysis. This continues in section 4 to derive useful closed-form expressions for the wealth, bequest amount, monetary consumption rate, and income rate (mortality credit rate). I define a monetary consumption-bequest ratio, MCBR. In section 5 the proportion of wealth invested in the tontine account is forced to be zero. Although this is generally sub-optimal, it corresponds to an investor following a pure decumulation strategy with bequest motive, and in the case of $b = 0$ or $\gamma \to 0$, also leads to closed-form solutions. Section 6 describes numerical results. Section 7 summarises and concludes.

## 2. Outline review of literature on pooled annuities and tontines

A person with a pension pot may choose to purchase an annuity from an insurance company, and enjoy a stream of income for the remainder of his/her life. The simplest form is a level annuity with a constant rate of payment. One may also purchase, at increased cost, an annuity with escalating income rate. Either way the purchaser is relieved of the risk associated with his/her own idiosyncratic and uncertain residual life span. This risk is transferred to the insurance company who will reduce it through the law of large numbers assuming a large enough number of annuitants. However, the insurance company also has to bear the systematic longevity risk arising from the fact that population mortality rates are not deterministic but are stochastic and difficult to predict. This will incur a risk premium that will be charged to annuitants. Other reasons why an annuity might appear to be poor value are the insurance company's need to cover administration costs, make a profit, and cover the cost of annuitants' adverse selection and moral hazard.

At the other extreme, an individual can choose to take personal responsibility for decumulation of a pension pot with all the inherent risks of running out of money in old age. However, this option does allow one more easily to satisfy a bequest motive and to follow a personal investment strategy. An example of attempting to combine such decumulation followed by annuitization at a later age in the presence of a bequest motive, is given by Gerrard et al. (2006).

A half-way house between these two alternatives is for a group of investors to collectively manage both idiosyncratic and systematic risks. These are pooled annuity, group self-annuitization, or tontine schemes. Piggot et al. (2005) design a group self-annuitization scheme. Members belonging to a cohort enter the scheme simultaneously with identical ages but bring heterogeneous wealth amounts to the pool. Each year the pay-out to surviving members is that calculated for an annuity based upon an agreed cohort lifetable. The pay-out is modified by a factor related to the ratio of actual to expected number of survivors. A pooled annuity scheme is examined by Stamos (2008) who considers a homogenous pool of annuitants who are identical copies in terms of entering age, risk aversion, initial wealth, and investment strategy. An actuarially fair distribution of a deceased's assets is simply to divide it equally between remaining pool members. There is no bequest motive. Sabin (2010) introduces a fair tontine annuity (FTA) tontine that allows heterogeneous membership with respect to age at entry, wealth, and investment strategy. Members can join at any time. On death of a member, each surviving member receives an explicit mortality credit that takes account of the surviving member's mortality rate and amount invested, in such a way that the expected income in any time period is the same as that pertaining to a fair annuity. Bequest motives are not considered. Donnelly et al. (2013) compare the pooled annuity of Stamos with a mortality-linked fund (essentially an annuity with freedom to invest in risky assets) showing that unless the fund costs are very small the expected return on the pooled annuity is better. When considering consumption in a life-cycle model, the expected utility was usually better for the pooled annuity.

The question of actuarial fairness of such schemes is discussed in Sabin (2010) and Donnelly (2015). The latter shows that the heterogeneous group self-annuitization scheme of Piggot et al. (2005) is neither actuarially fair or equitable. This amounts to some members, in expectation, subsidising others, with the extent amplified by the degree of heterogeneity. Sabin (2010) is shown to be actuarially fair for some but not all heterogeneous groups while the annuity overlay fund of Donnelly et al. (2014) is always actuarially fair due to the deceased member's estate receiving back a proportion of his/her own mortality credit to the group. Milevsky and Salisbury (2015) also consider a closed bequest-free implicit tontine comprising homogeneous membership. The group invests in a risk-free asset and a sponsor makes payments at a deterministic rate multiplied by the proportion of members who are still alive. As in all these schemes, systematic longevity risk is borne by the group. It appears that the scheme is equitable in that it treats all members equally in terms of expected lifetime-income. But despite the homogenous membership, it is not quite actuarially fair because there will be a small residual sum left over once the last member has died. In Milevsky and Salisbury (2016) the authors introduce heterogeneous membership. Bernhardt and Donnelly (2019) provide the focus for the present paper. The structure allows great flexibility. It features heterogeneous membership, with individualised investment strategies. On death of a member, explicit mortality credits are paid to surviving members in the manner described in section 1. It is actuarially fair, both instantaneously and over each member's lifetime, and is therefore also equitable, in that in expectation no member subsidises another. Crucially, it allows for a bequest motive, offering further flexibility over a conventional annuity. The purpose of this paper is to show how the bequest proportion may be optimised. In so doing it reveals a closed form and highly intuitive solution.

**3. Derivation of Optimal Consumption under instantaneous actuarial fairness**

Let $S(\boldsymbol{t})$ denote a finite set of members in such a tontine with current ages $\boldsymbol{t}$. Let $\alpha_i(t_i)$, $X_i(t_i)$, $\lambda_i(t_i)$ denote the tontine proportion, wealth, and mortality rate for member $i$ whose current age is $t_i$. Let $S_1(\boldsymbol{t}) = \{i \in S(\boldsymbol{t}) | \alpha_i(t_i) \geq 0\}$ and $S_2(\boldsymbol{t}) = \{i \in S(\boldsymbol{t}) | \alpha_i(t_i) < 0\}$ denote two dynamically changing subsets of $S(\boldsymbol{t})$. I propose the following rule. If $i \in S_1(\boldsymbol{t})$ and dies in the next time increment $dt$ then it generates a positive mortality credit for each $j \in S_1(\boldsymbol{t})$ of size $\alpha_i(t_i)X_i(t_i)\frac{\lambda_j(t_j)\alpha_j(t_j)X_j(t_j)}{\sum_{k\in S_1(t)}\lambda_k(t_k)\alpha_k(t_k)X_k(t_k)}$. Since this happens with probability $\lambda_i(t_i)dt$, summing over all $i \in S_1(\boldsymbol{t})$ leads in the limit of infinite $|S_1(\boldsymbol{t})|$ to a deterministic mortality credit rate to $j$ of $\lambda_j(t_j)\alpha_j(t_j)X_j(t_j)$, as shown in (3). Since this is also the monetary rate of $j$ donating mortality credits through her own death, instantaneous actuarial fairness is established. Conversely, if $i \in S_2(\boldsymbol{t})$ dies, then the rule is that $i$ would receive mortality credits from each $j \in S_2(\boldsymbol{t})$, of size $-\alpha_i(t_i)X_i(t_i)\frac{\lambda_j(t_j)\alpha_j(t_j)X_j(t_j)}{\sum_{k\in S_2(t)}\lambda_k(t_k)\alpha_k(t_k)X_k(t_k)}$, meaning that the terminal wealth of $i$ has instantaneously increased from $X_i(t_i)$ to $X_i(t_{i_+}) = (1-\alpha_i(t_i))X_i(t_i)$, all of which is in $i's$ bequest account.[1] Actuarial fairness follows in the same way as before. Note that these rules mean that mortality credits occur within but not between subsets and that when a member dies, he also donates/receives his share of the entire mortality credit associated with that death. The scheme guarantees instantaneous actuarial fairness in the same way as in Donnelly et al. (2014).

Given the control problem as stated in (2) and (3), I define a value function for a member, similar to that in Bernhardt and Donnelly (2019), namely

$$V(t,x) = \max_{\{w(s),\alpha(s),c(s):s\geq t\}} \mathbb{E}\Big(\int_t^\infty \exp\left[-\int_t^s[\lambda(u)+\rho]du\right]\left(U[c(s)X(s)] + b\lambda(s)U[(1-\alpha(s))X(s)]\right)ds\Big|X(t)=x\Big) \tag{12}$$

where the pension pot value (wealth function) follows

$$\frac{dX(t)}{X(t)} = \big(r + (\mu-r)w(t) + \alpha(t)\lambda(t) - c(t)\big)dt + \sigma w(t)dW(t) \tag{13}$$

where $W(t)$ is the standard Wiener process and where the rate of mortality credits $\alpha(t)\lambda(t)X(t)$ will be positive or negative depending upon the sign of $\alpha(t)$. The control functions are constrained so that $w(s) \geq 0, 1 \geq \alpha(s), c(s) \geq 0$ for all $s \geq t$. The proportion of funds invested in the risky asset might exceed 1. I allow for this possibility by assuming, for theoretical expediency, that borrowing is possible at the risk-free rate $r$. The proportion of wealth invested in a tontine, $\alpha(t)$, may be negative. Replacing the utility (1) by $U(x) = (x^\gamma - 1)/\gamma$ in (2) will yield the same optimal control values. Since $\lim_{\gamma\to 0}\frac{x^\gamma-1}{\gamma} = \ln x$, we identify this as the utility function for this limiting case.

A candidate optimal strategy is obtained using a dynamic programming Hamilton-Jacobi-Bellman approach, by solving

---

[1] There may be a practical difficulty for finite sized $S_2(\boldsymbol{t})$. Suppose $i$ dies. Then $X_j(t_{j_+}) = X_j(t_j) + \alpha_i(t_i)X_i(t_i)\frac{\lambda_j(t_j)\alpha_j(t_j)X_j(t_j)}{\sum_{k\in S_2(t)}\lambda_k(t_k)\alpha_k(t_k)X_k(t_k)}$. We must have $X_j(t_{j_+}) > 0$. Let $p_j(t) = \frac{\lambda_j(t_j)\alpha_j(t_j)X_j(t_j)}{\sum_{k\in S_2(t)}\lambda_k(t_k)\alpha_k(t_k)X_k(t_k)}$. Then we must have $p_j(t) < \frac{X_j(t_j)}{-\alpha_i(t_i)X_i(t_i)}$ for all $i,j \in S_2(\boldsymbol{t})$. Such a condition is more easily satisfied for large $|S_2(\boldsymbol{t})|$. For small $|S_2(\boldsymbol{t})|$ one might consider prescribing bounds on $\{X_j(t_j)\}$.

$$[\lambda(t)+\rho]V(t,x)=\max_{w,c,\alpha}\left[\frac{(cx)^\gamma}{\gamma}+b\lambda(t)\frac{[(1-\alpha)x]^\gamma}{\gamma}+\frac{\partial V(t,x)}{\partial t}+x\frac{\partial V(t,x)}{\partial x}[r+w(\mu-r)+\alpha\lambda(t)-c]+\frac{1}{2}x^2\sigma^2w^2\frac{\partial^2 V(t,x)}{\partial x^2}\right] \tag{14}$$

where I have now augmented the optimization to include the control $\alpha$. A proof of optimality for these candidate policies requires a verification result along the lines of the proof that is given in Bernhardt and Donnelly (2019). I will leave that for future research and here I will assume that the candidate policies derived below are indeed optimal and that $\lim_{t\to\infty}[1-\alpha^*(t)]=1$ if $b>0$.

Taking the first and second derivatives of the expression in brackets with respect to $c$ gives a maximum at

$$(c^*x)^{\gamma-1}x-x\frac{\partial V}{\partial x}=0 \tag{15}$$

which, with the hope that $c^*$ is independent of $x$, is suggestive of a trial solution of the form[2]

$$V(t,x)=\frac{{c^*}^{\gamma-1}x^\gamma}{\gamma}. \tag{16}$$

The first order condition on $w$ gives

$$x\frac{\partial V}{\partial x}(\mu-r)+wx^2\sigma^2\frac{\partial^2 V}{\partial x^2}=0. \tag{17}$$

From (16), $\frac{\partial^2 V}{\partial x^2}<0$ , which implies that the stationary point is a maximum, and that consequently

$$w^*(t)(\mu-r)=\left(\frac{1}{1-\gamma}\right)\left(\frac{\mu-r}{\sigma}\right)^2. \tag{18}$$

Note that $w^*(t)$ is independent of $t$ and so henceforth is expressed as $w^*$. Finally, the first order condition on $\alpha$ is

$$-b\lambda(t)x[(1-\alpha)x]^{\gamma-1}+\lambda(t)x\frac{\partial V(t,x)}{\partial x}=0. \tag{19}$$

And, differentiating again, it is seen that this is a maximum. Using (16) this gives

$$1-\alpha^*(t)=c^*(t)b^{\frac{1}{1-\gamma}}. \tag{20}$$

Substituting (16), (18), and (20) into (14)

$$[\lambda(t)+\rho]V(t,x)=c^*(t)V(t,x)+\lambda(t)c^*(t)V(t,x)b^{\frac{1}{1-\gamma}}+V(t,x)(\gamma-1)\frac{1}{c^*(t)}\frac{dc^*(t)}{dt}+\gamma V(t,x)[r+w^*(\mu-r)+\alpha(t)\lambda(t)-c^*(t)]+\frac{V(t,x)}{2}\gamma(\gamma-1)\sigma^2{w^*}^2, \tag{21}$$

[2] The approach differs from that of Bernhardt and Donnelly (2019) who suggest a value function $V(t,x)=h(t)x^\gamma$ and $c(t)=(\gamma h(t))^{\frac{1}{1-\gamma}}$. There appears to be an error in that these do not satisfy their Chini equation labelled equation (9) in their paper. However, if $c(t)=(\gamma h(t))^{\frac{1}{\gamma-1}}$ that does lead to their equation (9).

Recalling (5), that is

$$\frac{1}{c^*(t)}\frac{\mathrm{d}c^*}{\mathrm{d}t} = c^*(t)\left[1 + b^{\frac{1}{1-\gamma}}\lambda(t)\right] - \lambda(t) - \beta. \qquad (22)$$

Equation (22) is a Bernoulli differential equation with general solution

$$c^*(t) = \frac{1}{m(t) + \left[c^{*-1}(0) - m(0)\right]e^{\int_0^t [\lambda(u)+\beta]du}} \qquad (23)$$

where

$$m(t) = \int_t^\infty [1 + b^{\frac{1}{1-\gamma}}\lambda(u)]e^{-\int_t^u [\lambda(y)+\beta]dy}du. \qquad (24)$$

Now,

$$\begin{aligned} m(t) &= \int_t^\infty (1 + b^{\frac{1}{1-\gamma}}[\lambda(u) + \beta])e^{-\int_t^u [\lambda(y)+\beta]dy}du - \beta b^{\frac{1}{1-\gamma}}\int_t^\infty e^{-\int_t^u [\lambda(y)+\beta]dy}du \\ &= b^{\frac{1}{1-\gamma}} + \left[1 - \beta b^{\frac{1}{1-\gamma}}\right]A(t,\beta) \end{aligned} \qquad (25)$$

where

$$A(t,\beta) = \int_t^\infty e^{-\int_t^u [\lambda(y)+\beta]dy}du \qquad (26)$$

is the actuarially fair price of an annuity paying out at the rate of £1 per year from age $t$ to death, under a discount rate of $\beta$. Note that with an exponentially increasing mortality rate, $\lim_{t\to\infty} A(t,\beta) = 0$, and that $A(t,\beta)$ is decreasing in $t$.[3]

Let $s$ denote a member's age of entry to the tontine. Since $A(t,\beta)$ is decreasing in $t$, it follows from (25) that if $b^{\frac{1}{1-\gamma}}\beta < 1 + \frac{b^{\frac{1}{1-\gamma}}}{A(s,\beta)}$ then $m(t)$> 0 for all $t > s$. When $b > 0$, we restrict the search to those solutions that satisfy a boundary condition[4] that $\lim_{t\to\infty}[1 - \alpha^*(t)] = 1$. The reason is that with an exponentially increasing mortality rate, in the limit, an investor will heuristically place everything in the bequest account, since there is no opportunity to enjoy either consumption or mortality credits from the tontine account. Given an exponentially increasing mortality rate, if $c^{*-1}(0) - m(0) < 0$, then the denominator of (23) becomes negative for some $t$, violating $c^*(t) \geq 0$. If $c^{*-1}(0) - m(0) > 0$, then from (23), (25), and (26), $\lim_{t\to\infty} c^*(t) = 0$ leading to

3

$$\frac{\mathrm{d}A}{\mathrm{d}t} = -1 + [\lambda(t) + \beta]\int_t^\infty e^{-\int_t^u [\lambda(y)+\beta]dy}du < -1 + \int_t^\infty [\lambda(u) + \beta]\, e^{-\int_t^u [\lambda(y)+\beta]dy}du = -1 - e^{-\int_t^u [\lambda(y)+\beta]dy}\Big|_{u=t}^{\infty} = -e^{-\int_t^\infty [\lambda(y)+\beta]dy} < 0$$

4 For the case $b = 0$, the boundary condition is $\lim_{t\to\infty} c^*(t) = \infty$, which also leads to $c^{*-1}(0) = m(0)$.

$\lim_{t\to\infty}[1-\alpha^*(t)] = \lim_{t\to\infty} c^*(t) b^{\frac{1}{1-\gamma}} = 0$, violating the boundary condition. It follows that ${c^*}^{-1}(0) = m(0)$ and that

$$c^*(t) = \frac{1}{m(t)}. \tag{27}$$

Thus

$$c^*(t) = \frac{1}{b^{\frac{1}{1-\gamma}} + \left(1-\beta b^{\frac{1}{1-\gamma}}\right) A(t,\beta)} \tag{28}$$

and

$$1-\alpha^*(t) = c^*(t) b^{\frac{1}{1-\gamma}} = \frac{b^{\frac{1}{1-\gamma}}}{b^{\frac{1}{1-\gamma}} + \left[1-\beta b^{\frac{1}{1-\gamma}}\right] A(t,\beta)}. \tag{29}$$

$1/c^*(t)$ has a useful interpretation. It is the fair price of an annuity, subject to a discount rate of $\beta$, that pays out at the rate of £1 per year for life together with a terminal lump sum of £ $b^{\frac{1}{1-\gamma}}$ . We will show later that £ $b^{\frac{1}{1-\gamma}}$ is in fact an optimising tontine member's desired bequest : monetary consumption rate ratio.

From (28), as $t \to \infty$, $c^*(t) \to \frac{1}{b^{\frac{1}{1-\gamma}}}$ and $V(t,x) \to \frac{x^\gamma}{\gamma\left(b^{\frac{1}{1-\gamma}}\right)^{\gamma-1}} = \frac{bx^\gamma}{\gamma}$. When $\beta b^{\frac{1}{1-\gamma}} < 1$, using (29) we deduce that $\alpha^*(t) > 0$. An investor, who, as an alternative to tontine membership, practices pure decumulation of a pension pot, is essentially a member of an infinite sized tontine who sets $\alpha(t) = 0$, a sub-optimal choice. Therefore, actuarially speaking, a utility-maximising tontine member does better than such an investor. If $\beta b^{\frac{1}{1-\gamma}} = 1$, then $c^*(t) = b^{\frac{1}{\gamma-1}}$ and from (29), $\alpha^*(t) = 0$. That is, at all times, all of the pension pot is invested in the bequest account and the tontine member's utility is the same as an optimising decumulating investor. If the tontine member has a high enough bequest motive such that

$$1 < \beta b^{\frac{1}{1-\gamma}} < 1 + \frac{b^{\frac{1}{1-\gamma}}}{A(t,\beta)} \tag{30}$$

then (29) leads to $\alpha^*(t) < 0$ and negative mortality credits. The tontine member, say $j \in S_2(\boldsymbol{t})$, behaves rather like a life insuree, a person who buys life insurance. In this case the premiums are at a rate of $-\alpha_j^*(t_j)\lambda_j(t_j)$ of the member's current wealth, $X_j(t_j)$. On death of such a member at age $\tau$ say, the wealth of the deceased instantaneously increases by $-\alpha_j^*(\tau)X(\tau)$. This is funded by mortality credits made at that instant by all members of $S_2(\boldsymbol{t})$ . The deceased member's wealth is now $[1-\alpha^*(\tau)]X(\tau)$, all of which resides in the bequest account.

Rewriting (16) as $V(t,x) = \frac{{c^*}^\gamma(t)x^\gamma}{\gamma c^*(t)} = \frac{U(c^*(t)x)}{c^*(t)}$, we see that given (28), $|V(t,x)|$ is the actuarially fair price of an annuity purchased at age $t$, under a discount rate of $\beta$, that pays out at a constant rate of $|U(c^*(t)x)|$ together with a terminal lump sum of $b^{\frac{1}{1-\gamma}}|U(c^*(t)x)|$.

Note that with a logarithmic utility function, we set $\gamma = 0$, giving $\beta = \rho$. Thus when $\rho < b^{-1} + A(t,\rho)^{-1}$ we find that

$$c^*(t) = \frac{1}{b + [1 - b\rho\ ]A(t,\rho)} \tag{31}$$

which is identical to result (14) stated for $\rho < b^{-1}$ in Bernhardt and Donnelly (2019), who optimise for a fixed rather than dynamic $\alpha$, and dynamic $c(t)$. According to result (31), when $b^{-1} < \rho < b^{-1} + A(t,\rho)^{-1}$, the dynamic proportion invested in the bequest account is $bc^*(t) > 1$.

**4. Evolution of Wealth, Discounted Bequest Value and Discounted Monetary Consumption rate**

In this section, I now obtain a closed-form solution for the wealth equation under the optimal strategy. Using (20)

$$\begin{aligned}\frac{\mathrm{d}X(t)}{X(t)} &= \left(r + (\mu - r)w^* + \left[1 - c^*(t)b^{\frac{1}{1-\gamma}}\right]\lambda(t) - c^*(t)\right)dt + \sigma w^* dW(t)\\ &= \left(r + (\mu - r)w^* + \lambda(t) - c^*(t)\left[1 + b^{\frac{1}{1-\gamma}}\lambda(t)\right]\right)dt + \sigma w^* dW(t).\end{aligned} \tag{32}$$

But from (22)

$$c^*(t)\left[1 + b^{\frac{1}{1-\gamma}}\lambda(t)\right] = \frac{\mathrm{d}\ln[c^*(t)]}{\mathrm{d}t} + \beta + \lambda(t) \tag{33}$$

and so

$$\begin{aligned}\frac{\mathrm{d}X(t)}{X(t)} &= \left(r + (\mu - r)w^* + \lambda(t) - \frac{d\ln[c^*(t)]}{dt} - \lambda(t) - \beta\right)dt + \sigma w^* dW(t)\\ &= \left(r - \beta + (\mu - r)w^* - \frac{d\ln[c^*(t)]}{dt}\right)dt + \sigma w^* dW(t)\end{aligned} \tag{34}$$

Solving this subject to an initial condition that $X(s) = x_s$ we obtain for $t > s$

$$X(t) = \frac{x_s c^*(s) e^{\left(r - \beta + (\mu - r)w^* - 0.5\sigma^2 w^{*2}\right)(t-s)} e^{\sigma w^*[W(t) - W(s)]}}{c^*(t)} \tag{35}$$

and

$$\mathbb{E}\big(X(t)\big) = \frac{x_s c^*(s) e^{(r - \beta + (\mu - r)w^*)(t-s)}}{c^*(t)}\ . \tag{36}$$

The present value of the stochastic bequest amount, expressed as a proportion of the initial wealth, is given by

$$B(t) = \frac{e^{-r(t-s)}[1 - \alpha^*(t)]X(t)}{x_s} = b^{\frac{1}{1-\gamma}} c^*(s) e^{(-\beta + (\mu - r)w^*)(t-s)} e^{\sigma w^*[W(t) - W(s)] - 0.5\sigma^2 w^{*2}(t-s)} \tag{37}$$

while the corresponding present value of the stochastic monetary consumption rate is

$$C(t) = \frac{e^{-r(t-s)}c^*(t)X(t)}{x_s} = c^*(s)e^{(-\beta+(\mu-r)w^*)(t-s)}e^{\sigma w^*[W(t)-W(s)]-0.5\sigma^2 w^{*2}(t-s)} \quad . \tag{38}$$

If an investor wishes to aim for a level expectation of present value of monetary consumption and bequest amounts, we see that these are simultaneously achievable if

$$\beta = (\mu - r)w^*. \tag{39}$$

From (37) and (38) we see that the bequest amount on death is a fixed multiple, $b^{\frac{1}{1-\gamma}}$ of the monetary consumption rate at all ages. So, an investor who is not sure how to set $b$ can simply be asked what multiple of yearly consumption she would like to bequeath and so the model leads to a natural interpretation and setting of the bequest parameter through

$$b = \left(\frac{bequest\ amount}{monetary\ consumption\ rate}\right)^{risk\ aversion} \quad . \tag{40}$$

I now define a dimensionless monetary consumption-bequest ratio (MCBR) and rewrite (40) as

$$MCBR = b^{\frac{1}{\gamma-1}} \ . \tag{41}$$

From (29) the proportion invested in the bequest account is

$$1 - \alpha^*(t) = \frac{1}{1+(MCBR-\beta)A(t,\beta)} \quad . \tag{42}$$

While $1 + (MCBR - \beta)A(t,\beta) > 0$, given that $A(t,\beta)$ is decreasing in $t$ and that $\lim_{t\to\infty} A(t,\beta) = 0$ , it follows that $1 - \alpha^*(t)$ is increasing /decreasing in $t$ while $MCBR > \beta$ / $MCBR < \beta$, and that $\lim_{t\to\infty}(1 - \alpha^*(t)) = 1$. The condition for an investor to remain in $S_1(\boldsymbol{t})$ is that $MCBR > \beta$ and to remain in $S_2(\boldsymbol{t})$, it is $MCBR < \beta$. An investor may change her MCBR during a life-cycle in response to changing her level of risk aversion or bequest motive. A person with static MCBR is one who will use the tontine throughout the life cycle as providing annuity-style or life insurance-style benefits, but will not switch between the two.

The fractional consumption rate is

$$c^*(t) = [(1 - \alpha^*(t)]MCBR, \tag{43}$$

The expected value of the present value of the monetary consumption rate at time $t$, expressed as a proportion of initial wealth $x_s$ is

$$\mathbb{E}\big(C(t)\big) = \frac{c^*(t)x_s c^*(s)e^{(-\beta+(\mu-r)w^*)(t-s)}}{x_s c^*(t)} = c^*(s)e^{(-\beta+(\mu-r)w^*)(t-s)} \tag{44}$$

The expected value of the present value of the bequest amount expressed as a proportion of initial wealth is

$$\mathbb{E}\big(B(t)\big) = \frac{\mathbb{E}\big(C(t)\big)}{MCBR} \tag{45}$$

The expected value of the present value of the mortality credit rate (a positive or negative income rate) expressed as a proportion of initial wealth is

$$\mathbb{E}\big(I(t)\big) = \alpha^*(t)\lambda(t)\, e^{(-\beta+(\mu-r)w^*)(t-s)}\,\frac{c^*(s)}{c^*(t)} \tag{46}$$

## 5. The problem of pure decumulation with a bequest motive

It is relevant to consider the problem of pure decumulation (i.e., no tontine) of a pension pot under a bequest motive, by forcing $\alpha(t)$ to be zero. In that case

$$V(t,x) = \max_{\{w(s),c(s):s>t\}} \mathbb{E}\Big(\int_t^\infty \exp\big[-\int_t^s [\lambda(u)+\rho]du\big]\,\big(U[c(s)X(s)] + b\lambda(s)U[X(s)]\big)ds\Big| X(t)=x\Big) \tag{47}$$

subject to

$$\frac{\mathrm{d}X(t)}{X(t)} = \big(r + (\mu-r)w(t) - c(t)\big)dt + \sigma w(t)dW(t)\,. \tag{48}$$

The Hamilton-Jacobi-Bellman approach gives

$$[\lambda(t)+\rho]V(t,x) = \max_{w,c}\left[\frac{(cx)^\gamma}{\gamma} + b\lambda(t)\frac{x^\gamma}{\gamma} + \frac{\partial V(t,x)}{\partial t} + x\frac{\partial V(t,x)}{\partial x}[r + w(\mu-r) - c] + \frac{1}{2}x^2\sigma^2 w^2\frac{\partial^2 V(t,x)}{\partial x^2}\right]. \tag{49}$$

The first and second order conditions on $c$ and $w$ are as before, as is the expression for $V(t,x)$. Substituting these into (49)

$$\frac{1}{c^*(t)}\frac{\mathrm{d}c^*}{\mathrm{d}t} = c^*(t) + \frac{b\lambda(t)c^*(t)^{1-\gamma}}{1-\gamma} - \psi(t) \tag{50}$$

where

$$\psi(t) = \lambda(t) + \beta + \frac{\gamma\lambda(t)}{1-\gamma}. \tag{51}$$

Equation (50) is a Bernoulli differential equation only when $\gamma = 0$ or $b = 0$, so closed-form solutions are possible only in these cases. Stamos (2008) considers decumulation of pooled annuity funds with no bequest motive and has derived, for the degenerate case of only one member in the pool, an identical result to (50-51) for $b = 0$, his equation (32). Similarly, he has derived a closed-form solution, again for $b = 0$, his equation (34), for the special case of a pool of infinite size, which is the same as my tontine solution $c^*(t) = \frac{1}{A(t,\beta)}$ as in (28). When $\gamma = 0$, equations (50-51) do admit the closed-form solution (31) even when $b \neq 0$. It is exactly the same as the tontine solution, that is result (22) with $\psi(t) = \lambda(t) + \rho$. Therefore, we conclude that for a logarithmic utility with bequest motive, the optimal fractional consumption rate for pure decumulation and tontine are the same. That is to be expected as it can be seen from the objective function that the optimization for the consumption rate is the same in both cases. Further, subtracting $\gamma^{-1}$ from the utility function and taking the limit as $\gamma \to 0$ gives the expected lifetime utility as $V(t,x) = \frac{\ln[xc^*(t)]}{c^*(t)}$.

**6. Numerical Examples**

Some numerical results are shown for a Gompertz-Makeham mortality rate

$$\lambda(t) = v + \frac{e^{\frac{t-m}{q}}}{q} \tag{52}$$

for a UK male, where $m = 83.43, q = 10.94, v = -0.0052$, fitted to UK Office of National Statistics Life tables, for those aged 50 or more. We consider a member who enters the tontine at age $s = 65$.

Figure 1 shows how the bequest proportion changes as MCBR increases, for a range of $\beta$ values. Thus, an investor who sets MCBR to 0.01 wishes his bequest amount to be 100 times his yearly monetary consumption rate. For $\beta > MCBR$, the member is behaving as a life insuree. The mortality credits are negative and represent pseudo life insurance premiums. Expressed as a proportion of his wealth, they are the product of an increasing mortality rate and decreasing $|\alpha^*(t)|$ . When $\beta = MCBR$, he neither receives nor makes mortality payments as $\alpha^*(t)$ =0. The entire pension pot is always invested in the bequest account and he is like an independent investor who is decumulating the pot with a constant fractional consumption rate. When $\beta < MCBR$, the member is a pseudo annuitant, receiving mortality credits (income), which again as proportion of wealth, are the product of an increasing mortality rate and decreasing $\alpha^*(t)$. When MCBR=0.1 the investor wishes the bequest amount to be 10 times the monetary consumption rate. For specified risk aversion the bequest motive is now not so strong. As a result, for the values of $\beta$ shown, the member behaves as an annuitant. For large $\beta$ there is always a significant amount invested in the bequest account, increasing with age. But for low $\beta$ the initial bequest proportion is low. When MCBR=1, the investor has virtually no bequest motive, requiring a bequest to be the same as yearly monetary consumption. The member is again a pseudo annuitant and at lower ages places a small but increasing proportion in the bequest amount, moderately independent of $\beta$.

Figure 1. Optimal bequest proportion as a function of MCBR and $\beta$

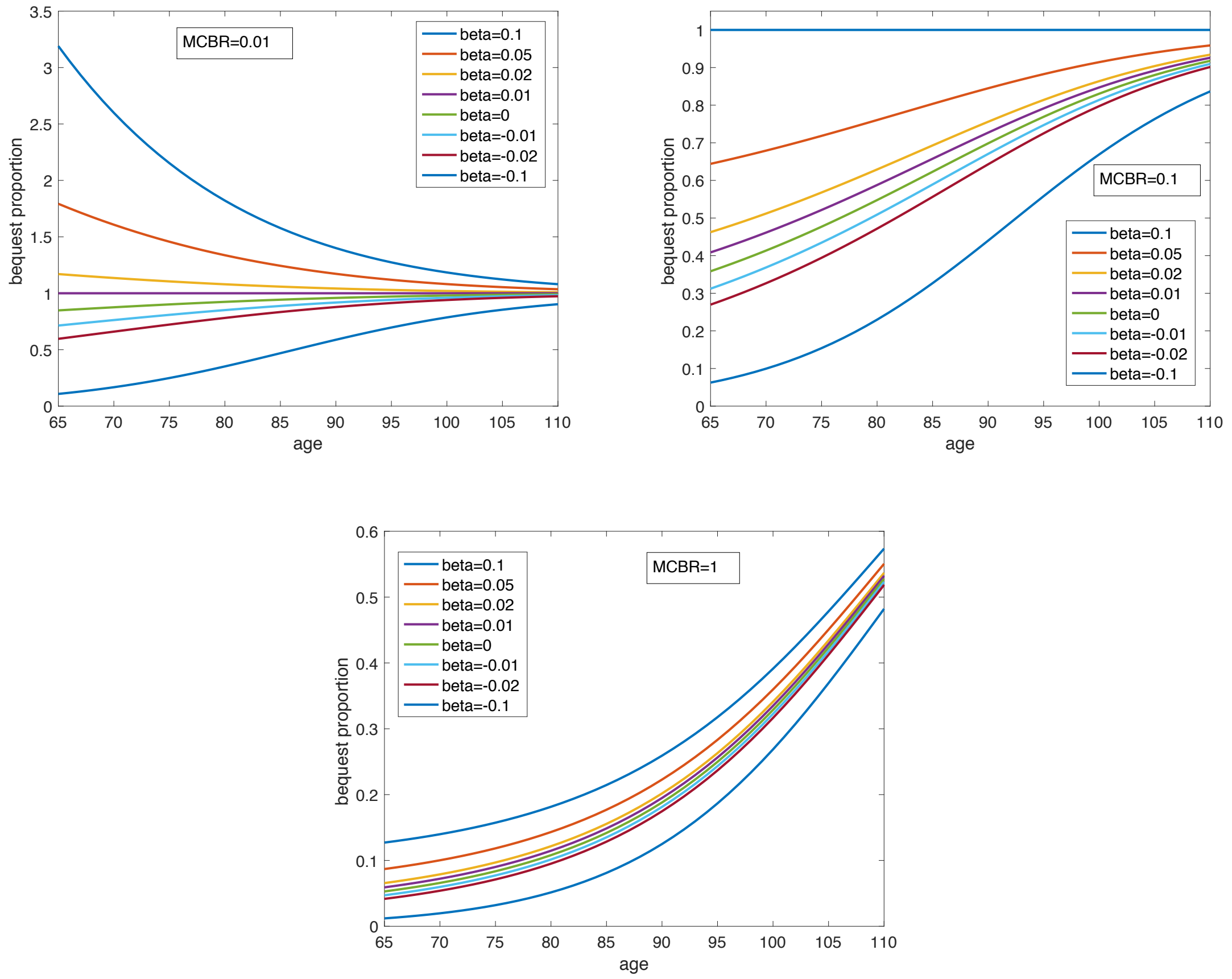

In Figures 2-5, I take $r = \rho = 0.02, \mu = 0.05, \sigma = 0.2$, and examine behaviour as $\gamma$ and $b$ change for a male who starts decumulating their pension pot at age $s = 65$ years. All monetary values are expectations and are present-valued to age 65 and expressed as a proportion of initial wealth. Figure 2 shows the bequest proportion, figure 3 the monetary consumption per year, figure 4 the bequest value, and figure 5 the mortality credit rate. I examine bequest motives in the range $b \in [0,60]$ and constant relative risk aversions in the range $1 - \gamma \in (0,11)$. With these values the proportion of current wealth invested in the risky asset is $w^* = \frac{0.05-0.02}{0.2^2(1-\gamma))}$ and so this does exceed 1 when $\gamma > 0.25$. In that case, borrowing to allow over investment in the risky asset raises the probability of running down the pension at an early age, alongside a small probability of large wealth at an advanced age. These aspects are consistent with a low relative risk aversion, $1 - \gamma$.

For $\gamma = 0.8$, $w^* = 3.75$ representing large borrowing to invest in the risky asset. In the early years the bequest proportion and consumption are small in order to gain mortality credits which are invested in the risky asset. It is only in old age that significantly more is transferred into the bequest account. This again is consistent with very low risk aversion and can result in huge bequest amounts, albeit with infinitesimally small probabilities. Figure 6 shows an extreme example of the highly skewed distribution of bequest amount at age 95 where $b = 3$. The expected bequest amount is 89. However, the most likely bequest is essentially zero, the median 0.02, and there is a 95% chance that the bequest amount is less than 17.

For $\gamma = 0.25$, $w^* = 1$ and all the wealth is invested in the risky asset. For very high bequest motives the member behaves as a life insuree with steadily increasing life insurance premiums and bequest amounts (terminal pay-out on death) but low consumption. For lower bequest motives, consumption is increasing with age but decreasing in bequest motive. Bequest amounts increase with age and with bequest motive.

For $\gamma = -0.08225$ with very high bequest motive almost all of the wealth is invested in the bequest account. Accordingly, income in the way of mortality credits is always very low as is consumption. This value of $\gamma$ gives level expected and present valued bequest and monetary consumption as in (39), but the actual outcomes are highly variable depending upon risky asset performance. Figure 6 shows an example of the log-normal distribution of present value of bequest amount at age 95 when $b = 3$. The expected bequest is 0.17, the median 0.13, and the mode 0.07.

For $\gamma = -10$ risk aversion is very high resulting in almost all of the wealth being invested risk-free, with low expected bequest amounts. Unlike the other less risk averse scenarios, we see a gradual decrease with age in expected monetary consumption.

Figure 2. Optimal bequest proportion as a function of $b$ and $\gamma$ for the case $r = \rho = 0.02, \mu = 0.05, \sigma = 0.2$

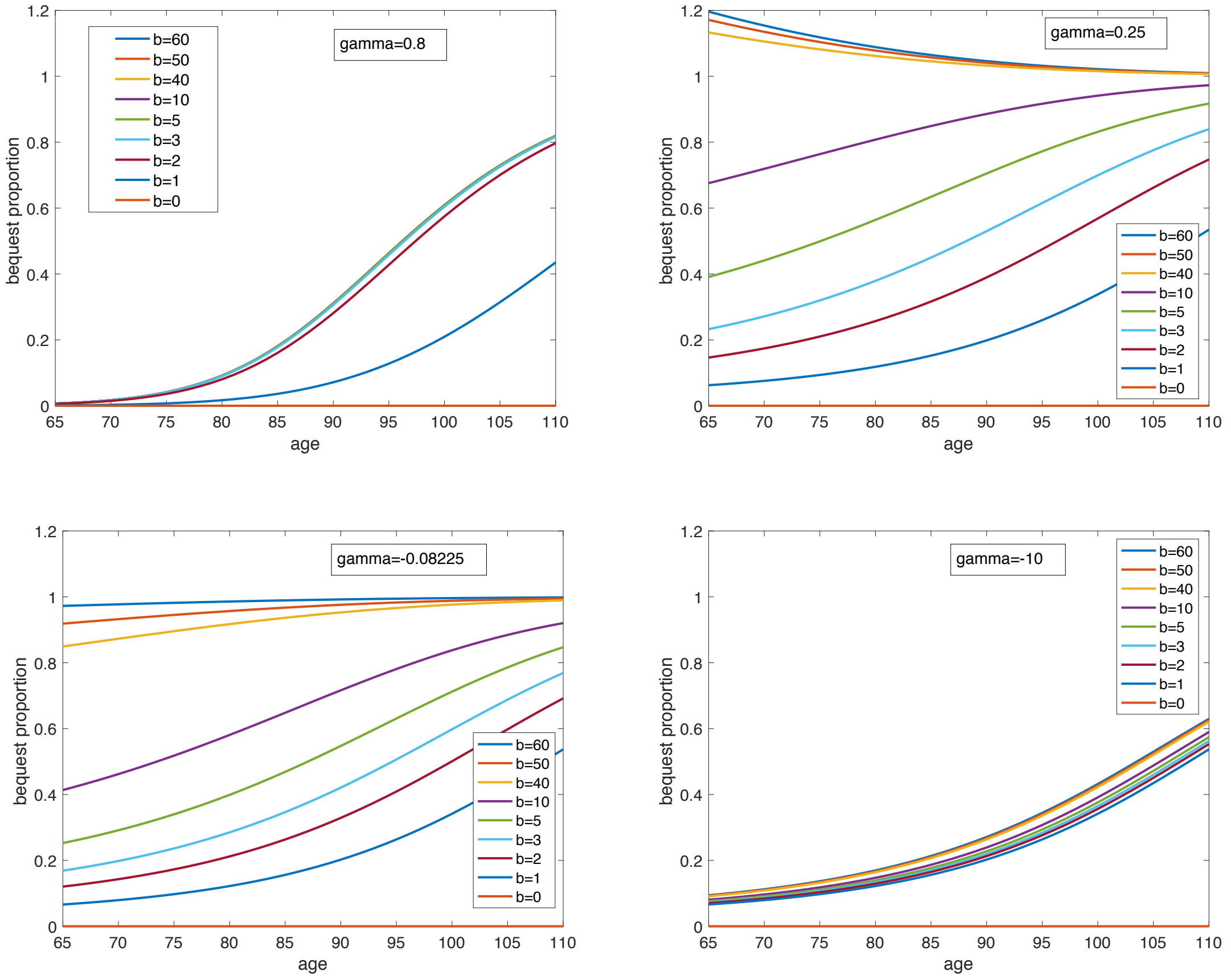

Figure 3. Expectation of present value of monetary consumption rate as a proportion of wealth at age 65 for $r = \rho = 0.02, \mu = 0.05, \sigma = 0.2$

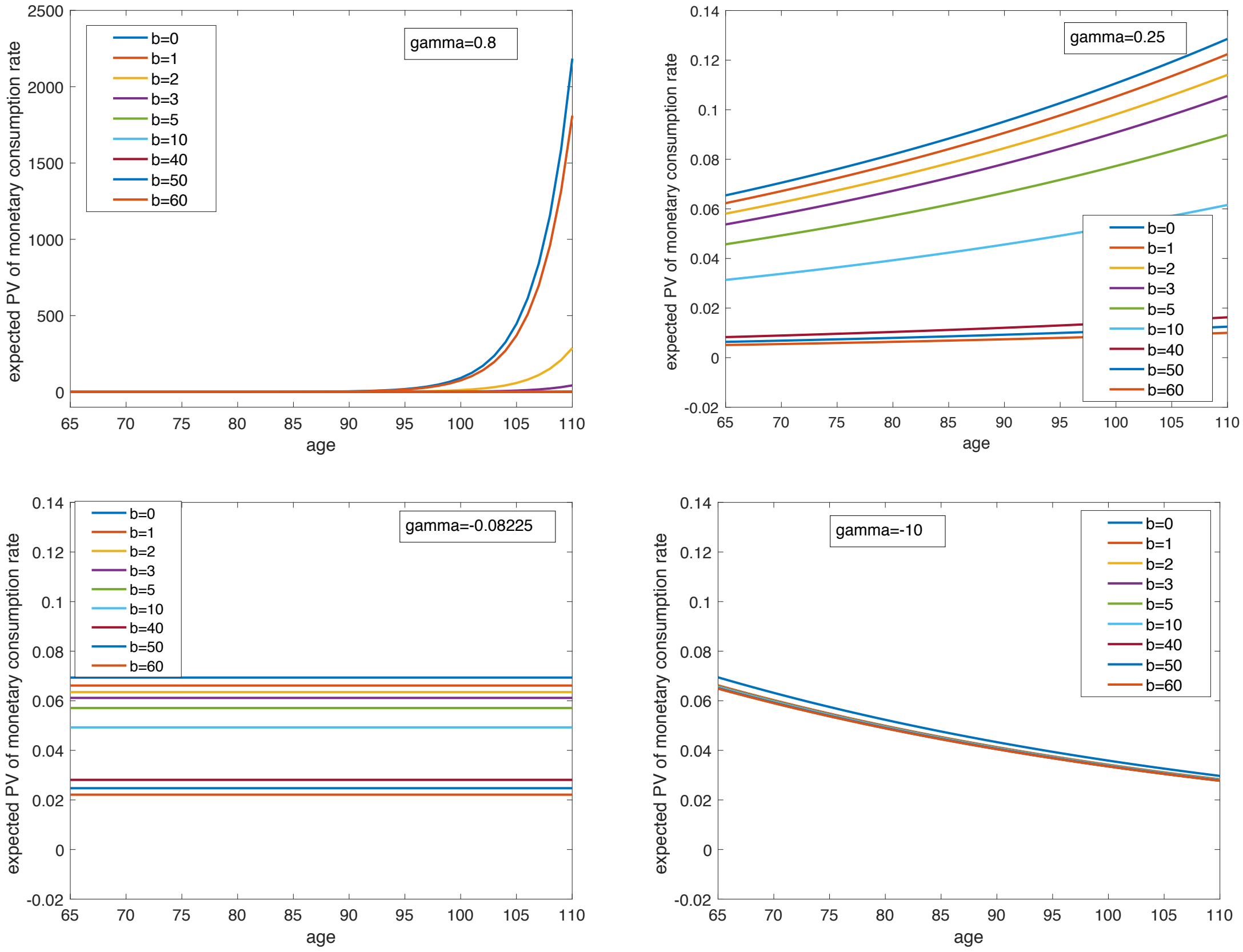

Figure 4. Expectation of present value of bequest amount as a proportion of wealth at age 65 for $r = \rho = 0.02, \mu = 0.05, \sigma = 0.2$

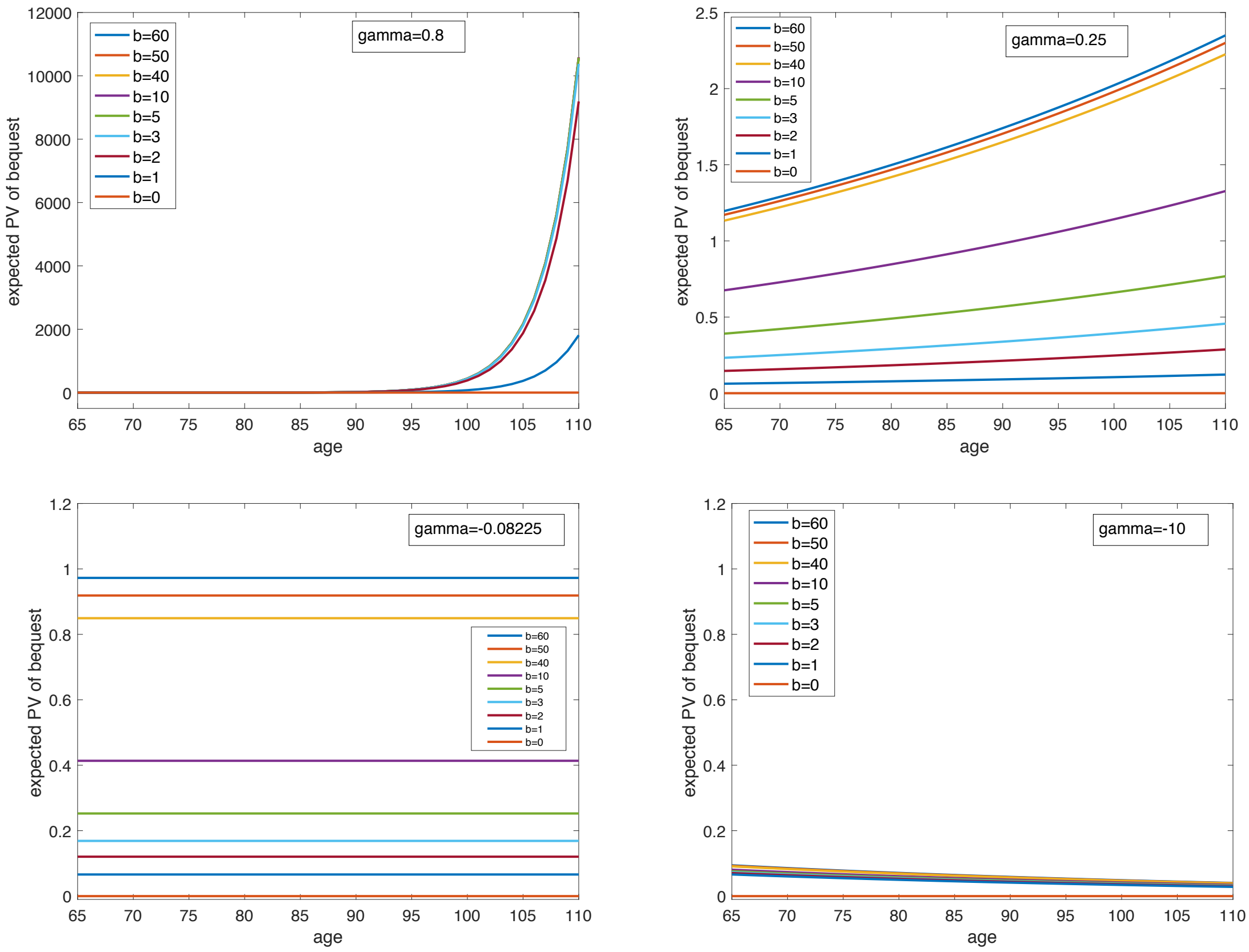

Figure 5. Expectation of present value of mortality credit rate as a proportion of wealth at age 65 for $r = \rho = 0.02, \mu = 0.05, \sigma = 0.2$

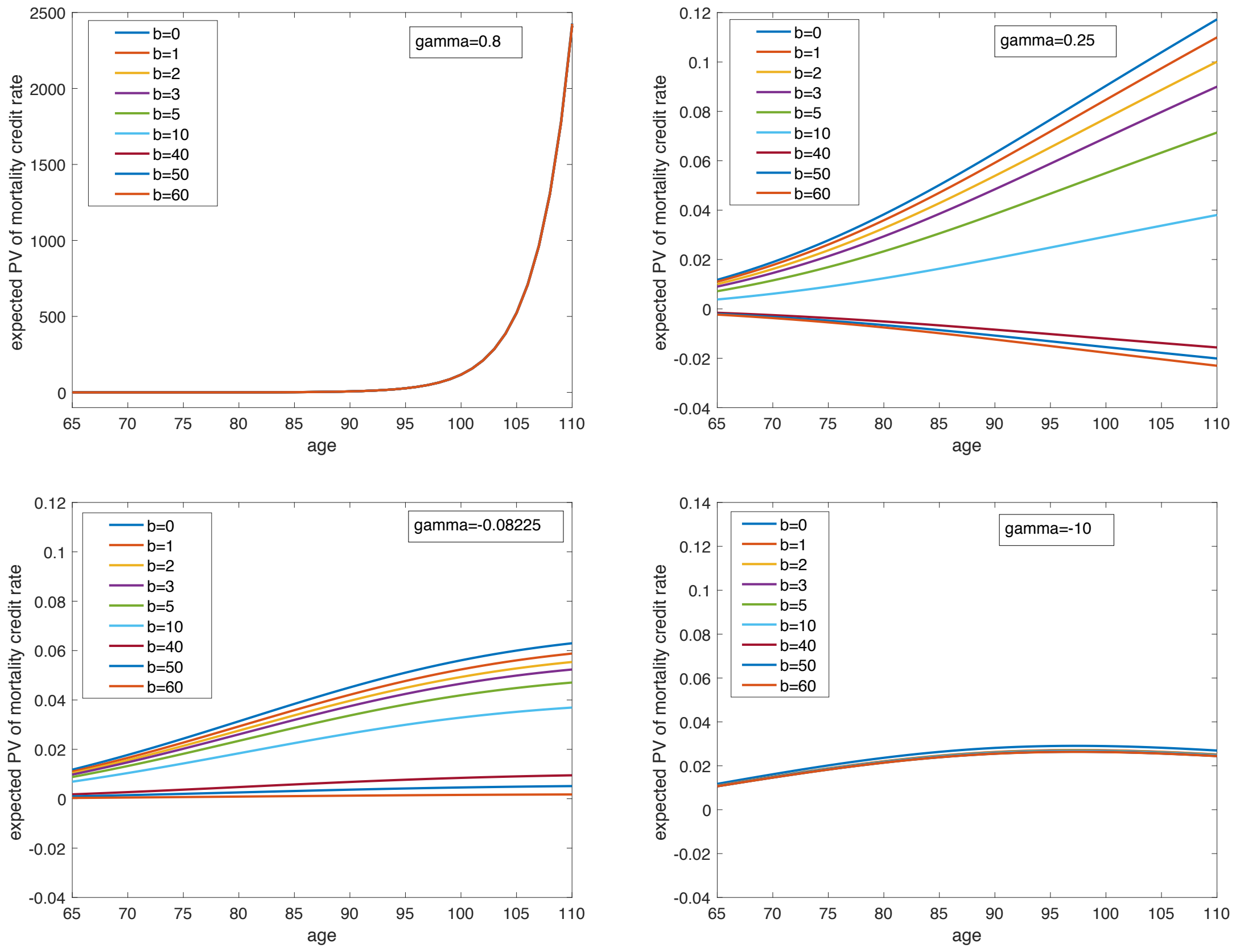

Figure 6. The variability of the present value of the bequest amount at age 95 as a proportion of the wealth at age 65 for $b = 3$ and $\gamma = -0.08225$ and 0.8

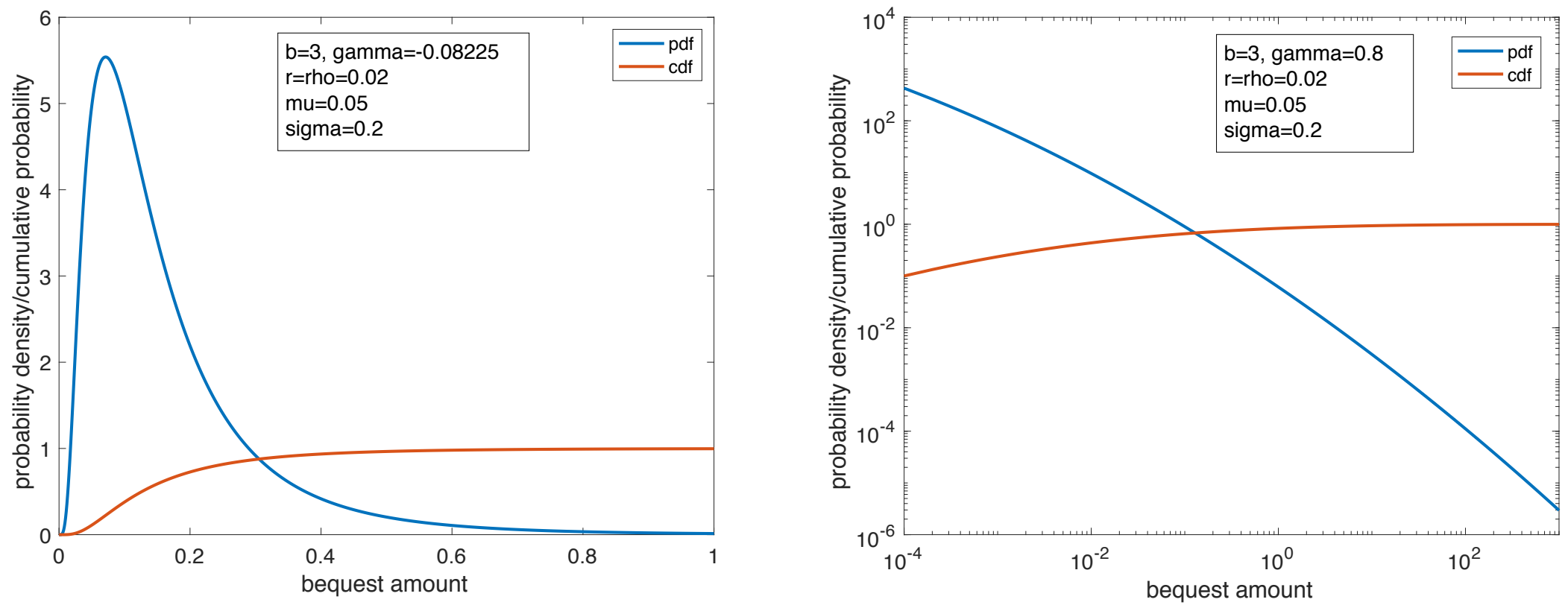

## 7. Summary and Conclusions

In this paper I have obtained closed-form solutions for the optimal fractional consumption rate, bequest-proportion, wealth, and risky/riskless investment ratio for a modern, ideal, explicit tontine with bequest motive, under a constant relative risk aversion utility function and under the assumption that the corresponding verification result can be proven. It is shown that the first two are functions of the individual's age, independently of the wealth. The optimal bequest-proportion is a product of the optimal fractional consumption rate and an exponentiated bequest parameter. The reciprocal of the optimal fractional consumption rate turns out to be the fair price of an annuity, subject to a discount rate of $\beta$, that pays out at the rate of £1 per year for life together with a terminal lump sum of £ $b^{\frac{1}{1-\gamma}}$. Some numerical results are given based upon a mortality rate fitted to UK Office of National Statistics life tables.

How should an investor decide upon appropriate values for $1-\gamma$ (risk aversion) and $b$ (strength of bequest motive)? The former could be set using $1-\gamma = \frac{\mu - r}{w^* \sigma^2}$, where $w^*$ is the desired proportion in the risky asset. Alternatively, if the objective is to achieve level expected discounted monetary consumption rates and bequest amounts, then one would resort to result (39). The model leads to a natural interpretation and setting of the bequest parameter through

$$b = \left(\frac{bequest\ amount}{monetary\ consumption\ rate}\right)^{risk\ aversion}. \tag{53}$$

It is also of interest to note that a related problem of how to decumulate a pension pot (no tontine) under a bequest motive, is solvable by suppressing the optimization with respect to the bequest proportion, in favour of forcing it to be 1. In this case, closed-form solutions for the fractional consumption rate and wealth are now available for the following: all bequest motives with a logarithmic utility function; all constant relative risk aversion utilities where there is no bequest motive. With a logarithmic utility, the optimal fractional consumption rate is identical to that obtained for the corresponding tontine.

It is noticeable that for the case where the tontine member receives mortality credits $(\alpha^*(t) > 0)$, then the bequest proportion increases with age. That might appear to disproportionately disadvantage those who die early to the benefit of those who live to a good age. That is also a feature of annuities, although the latter have absolutely no bequest feature. Future research can explore whether strategies other than pure utility maximization are more in line with intuitive expectations of investors. Suffice to say that in the example given, this increase in bequest proportion is necessary to maintain, for example, a desire for a level present value of expected monetary consumption rate and monetary bequest amount, and that for someone wishing to escape from the drawbacks of an annuity, that might seem a reasonable objective. Thus, for example, we see from figures 3 and 4 that, when $\gamma = -0.0825$ and $b = 10$, the expectation of the present value of monetary consumption rate is at all times around 5 % of the value of the initial pension pot and the present value of the bequest amount is a multiple 8.4 of that, that is 42% of the initial pension pot. For other scenarios, we note that the increasing bequest proportion feature appears to result from the finding that the monetary consumption rate should at all ages be a fixed multiple of the bequest amount. However, with increasing age, an investor might feel it is more intuitive to increase that multiple. In essence, that means that the investor prefers to specify a decreasing bequest profile $b(t)$, instead of a constant $b$. Future research can explore such a variation of the model presented here.

I do not consider some of the practical problems that will need attention in implementing such a scheme. One such issue is that of adverse selection. For example, a person suffering a sharp reduction in life expectancy might opt to shift a large amount to the bequest account which would compromise the actuarial fairness of the scheme. A second example is that the behaviour of a finite sized tontine (rather than the infinite sized one addressed here) will be somewhat different and introduce more risk to members. This could be explored through simulation.

Regarding the dynamic changing of bequest proportion, it is noted that that this does not involve physical movements of assets, merely a formal declaration as to a member's current allocation. In fact, under the stated assumptions, $\alpha^*(t)$ is dynamic, but deterministic. It could therefore be prescribed in perpetuity when a member enters the tontine. However, that would remove some flexibility (and freedom to behave sub-optimally – which does not jeopardise actuarial fairness) from each member. In practice, it might be preferable to retain such flexibility, where, subject to managing adverse selection, actuarial fairness prevails.

A new finding is that for very large bequest motives, mortality credits are negative and as a result a tontine member behaves as a life insuree. For a finite sized tontine, it might be necessary to prescribe limits on such individuals' wealth to avoid the possibility of exhausting the pension pot.

A referee has suggested that the structure of the results will carry over if the main parameter values are deterministically time varying. Conjectured results are shown without proof in Appendix 1.

Bernhardt and Donnelly's (2019) paper breaks new ground in the theory of tontines. It is of practical importance given a growing interest in alternatives to both annuities and pure decumulation of pension pots. Compared with many pooled annuity schemes it has the considerable advantage of preserving instantaneous actuarial fairness under heterogeneous membership and random entry to the tontine. That paper does not consider a dynamic optimization of the bequest proportion and it transpires, fortuitously, that doing so simplifies the analysis and form of the solutions, and reveals the life insurance aspect of such a tontine.

**Acknowledgements**

The author thanks two anonymous referees who made suggestions on an earlier manuscript which resulted in a much improved paper. He also thanks Mr Jingkai Hong for fitting the parameter values in equation (52).

**Appendix 1**

Let $s$ denote age of entry in tontine. Let

$$\beta(t) = r(t) + \frac{\rho(t)-r(t)}{1-\gamma} - \frac{\gamma}{2}\left(\frac{\mu(t)-r(t)}{\sigma(t)}\right)^2 \left(\frac{1}{1-\gamma}\right)^2 \tag{54}$$

If for all $t \geq s$

$$1 + b^{\frac{1}{\gamma-1}} \int_t^\infty e^{-\int_t^u [\lambda(y)+\beta(y)]dy}\, du \geq \int_t^\infty \beta(u) e^{-\int_t^u [\lambda(y)+\beta(y)]dy}\, du \tag{55}$$

then

$$w^*(t) = \frac{\mu(t)-r(t)}{[1-\gamma]\sigma^2(t)} \tag{56}$$

$$c^*(t) = \frac{1}{b^{\frac{1}{1-\gamma}} + \int_t^\infty \left[1 - b^{\frac{1}{1-\gamma}}\beta(u)\right] e^{-\int_t^u [\lambda(y)+\beta(y)]dy} du} \tag{57}$$

$$1 - \alpha^*(t) = c^*(t) b^{\frac{1}{1-\gamma}} \tag{58}$$

$$X(t) = \frac{x_s c^*(s) e^{\int_s^t [r(u)-\beta(u)+(\mu(u)-r(u))w^*(u)-0.5\sigma^2(u){w^*}^2(u)]du + \int_s^t w^*(u)\sigma(u)dW(u)}}{c^*(t)} \tag{59}$$